\documentclass[a4paper,fleqn,usenatbib]{mnras}

\usepackage[T1]{fontenc}
\usepackage{ae,aecompl}
\usepackage{verbatim}

\usepackage{graphicx}	
\usepackage{amsmath}	
\usepackage{amssymb}	
\usepackage{multirow}

\title[D/H in subDLA system towards J\,1444$+$2919]{The primordial deuterium abundance:\\
subDLA system at $z_{\rm abs}=2.437$ towards the QSO J\,1444$+$2919}

\author[S. A. Balashev et al.]{
S. A. Balashev,$^{1,2}$\thanks{E-mail: s.balashev@gmail.com}
E. O. Zavarygin,$^{1,2}$
A. V. Ivanchik,$^{1,2}$
K. N. Telikova$^{2}$ and
\newauthor D. A. Varshalovich$^{1,2}$
\\
$^{1}$Ioffe Institute, Polytekhnicheskaya 26, Saint-Petersburg, 194021, Russia\\
$^{2}$Peter the Great St.-Petersburg Polytechnic University, Polytekhnicheskaya 29, Saint-Petersburg, 195251, Russia
}

\pubyear{2015}

\begin{document}
\label{firstpage}
\pagerange{\pageref{firstpage}--\pageref{lastpage}}
\maketitle

\begin{abstract}
We report a new detection of neutral deuterium in the sub Damped Lyman Alpha system with low metallicity [O/H]\,=\,$-2.042 \pm 0.005$ at $z_{\rm abs}=2.437$ towards QSO~J\,1444$+$2919. The hydrogen column density in this system is log$N$(H\,{\sc i})~$=19.983\pm0.010$ and the measured value of deuterium abundance is log(D/H)~$=-4.706\pm0.007_{\rm stat}\pm0.067_{\rm syst}$. This system meets the set of strict selection criteria stated recently by Cooke et al. and, therefore, widens the {\it Precision Sample} of D/H. However, possible underestimation of systematic errors can bring bias into the mean D/H value (especially if use a weighted mean). Hence, it might be reasonable to relax these selection criteria and, thus, increase the number of acceptable absorption systems with measured D/H values. In addition, an unweighted mean value might be more appropriate to describe the primordial deuterium abundance. The unweighted mean value of the whole D/H data sample available to date (15 measurements) gives a conservative value of the primordial deuterium abundance (D/H)$_{\rm p}=(2.55\pm 0.19)\times10^{-5}$ which is in good agreement with the prediction of analysis of the cosmic microwave background radiation for the standard Big Bang nucleosynthesis. By means of the derived (D/H)$_{\rm p}$ value the baryon density of the Universe $\Omega_{\rm b}h^2=0.0222\pm0.0013$ and the baryon-to-photon ratio $\eta_{10} = 6.09\pm 0.36$ have been deduced. These values have confident intervals which are less stringent than that obtained for the {\it Precision Sample} and, thus, leave a broader window for new physics. The latter is particularly important in the light of the lithium problem.
\end{abstract}



\section{Introduction}

The standard Big Bang nucleosynthesis (BBN) theory predicts abundances of light nuclei such as H, D, $^3$He, $^4$He and $^7$Li as a function of the baryon-to-photon ratio $\eta\equiv n_{\rm b}/n_{\gamma}$ \citep[e.$\,$g.,][]{Weinberg2008,Gorbunov2011}. A measurement of a ratio of any two primordial abundances determines $\eta$ and, hence, the baryon density $\Omega_{\rm b}$ as $\eta_{10}=10^{10}\eta \simeq 273.9\Omega_{\rm b} h^2$ \citep{Steigman2006} where $h$ is the dimensionless Hubble parameter. As a rule, hydrogen is used as one of these two elements and another one is determined with respect to H. Once $\eta$ is known, the primordial abundances of all the other elements are predicted and their measurements can be used to test extensions of the $\Lambda$CDM model (e.\,g., post-BBN decays of massive particles, \citealt{Jedamzik2004,Kawasaki2005,Pospelov2010}) and the Standard Model of physics (e.\,g., addition neutrino species, \citealt{Steigman2012}). Given its sensitivity to and monotonic dependence on $\eta$, the primordial abundance of deuterium (namely D/H ratio) is generally accepted as the best ``baryometer'' among all the aforementioned elements.

The deuterium abundance can be determined from absorption features in spectra of distant quasars (QSOs), namely, by means of D\,{\sc i} and H\,{\sc i} absorption lines \citep[e.$\,$g.,][]{Burles1998a, Burles1998b, OMeara2001, Kirkman2003}. In addition, complementary HD/2H$_2$ technique was suggested \citep{Balashev2010,Ivanchik2010,Ivanchik2015}, however, due to complexity of chemistry such estimations face some difficulties \citep{Liszt2015}. To date, there have been 14 absorption systems with estimations of the primordial D/H value based on D\,{\sc i}/H\,{\sc i} lines in QSO absorption systems (see Section~\ref{sec:D_problem}). However, while the mean D/H value is in reasonable agreement with BBN prediction, the individual measurements of the D/H value show pronounced scatter which is noticeably larger than the published errors. Given the important role of the deuterium in our understanding of standard BBN and its extensions, it is vital to widen the sample of measurements of the primordial D/H value.

In this paper we report a new (15th) measurement of the D/H value in the sub-damped Ly$\,\alpha$ (subDLA) at $z_{\rm abs}=2.437$ towards the QSO J\,1444$+$2919 as well as discuss the current situation with measurements of the primordial deuterium abundance in QSO absorption systems.

\section{Data}

The quasar J\,1444$+$2919 was observed at the Keck telescope using the High Resolution Echelle Spectrograph (HIRES, \citealt{Vogt1994}) during several independent programs. The data obtained with the last version of HIRES only have been used, namely in 
2007 (PI: Sargent, program ID: C203Hb) and in 2009 (PI: Steidel, program ID: C168Hb). The journal of the observations and of the exposure specifications is given in Table~\ref{tab:obs}. The data have been downloaded from the Keck Observatory Archive (KOA)\footnote{\url{https://koa.ipac.caltech.edu/cgi-bin/KOA/nph-KOAlogin}}. The C1 and C5 deckers have widths of slits of 0.861 and 1.148, respectively, resulting in resolution of 48000 (the instrument function width of 6.2 km/s) and 36000 (8.3 km/s), respectively.

For the data reduction, the MAKEE package\footnote{\url{http://www.astro.caltech.edu/~tb/ipac_staff/tab/makee/index.html}} developed by Tom Barlow has been used. The spectral reduction has been done in a usual way, using the calibration files for each exposure provided by the Keck archive.
The standard ThAr lamps have been used for the wavelength calibration of the exposures. For each exposure the nearest arc file have been taken. Since it is well known that echelle spectrum calibration at HIRES/KECK is not stable, a cross-correlation correction has been applied before coadding the exposures (see Section~\ref{subs:wl_cal}). The exposures with the same decker setups (i.e. resolution) have been combined with each other, resulting in two spectra, which have jointly been used for the analysis of the absorption system. The final spectra have very high Signal to Noise (S/N) ratios: the average values are $\sim60$ and $\sim50$ for C1 and C5 decker setups, respectively. In the following, the spectrum with 48000 resolution only is used for visualization, while all the estimates of the parameters will be made using both spectra jointly.

\begin{table*}
	\begin{center}
		\begin{tabular}{|c|c|c|c|c|c|c|}
			\hline
			No. & Date & Starting time, UT & decker  & Exposure, sec & Wavelenghts, \AA & Airmass\\
			\hline
			1 & 04.08.2007 & 13:11:48 & C1 & 3600  & 3130\,...\,6050 &  1.05\\
			2 & 04.08.2007 & 14:12:47 & C1 & 3600  & 3130\,...\,6050 &  1.17 \\
			3 & 04.09.2007 & 12:05:29 & C1 & 3600  & 3130\,...\,6050 &  1.01 \\
			4 & 04.09.2007 & 13:06:21 & C1 & 3600  & 3130\,...\,6050 &  1.05 \\
			5 & 25.05.2009 & 08:21:43 & C5 & 2000  & 2970\,...\,5940 &  1.02 \\
			6 & 25.05.2009 & 08:55:56 & C5 & 2000  & 2970\,...\,5940 &  1.01 \\
			7 & 25.05.2009 & 09:35:10 & C5 & 2000  & 2970\,...\,5940 &  1.03 \\
			8 & 25.05.2009 & 10:09:23 & C5 & 2000  & 2970\,...\,5940 &  1.06 \\
			9 & 25.05.2009 & 10:43:36 & C5 & 2000  & 2970\,...\,5940 &  1.12 \\
			\hline
			total & \multicolumn{3}{c}{} & 24400 \\
			\hline
		\end{tabular}
		\caption{Journal of the observations.}
		\label{tab:obs}
	\end{center}
\end{table*}

\subsection{Wavelength calibration}
\label{subs:wl_cal}

\begin{figure*}
	\centering
	\begin{tabular}{cc}
		\includegraphics[clip=,width=\columnwidth]{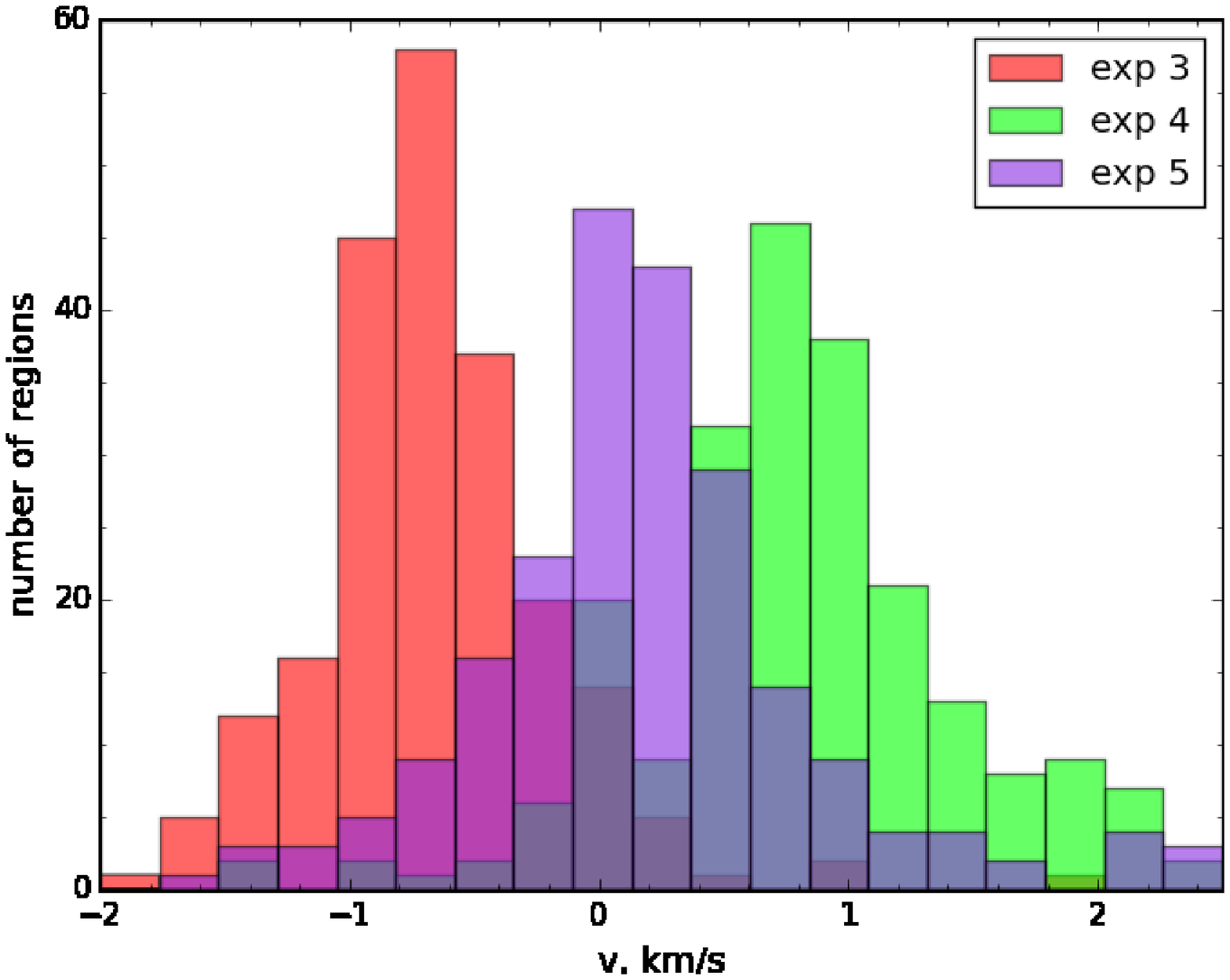} & 
		\includegraphics[clip=,width=\columnwidth]{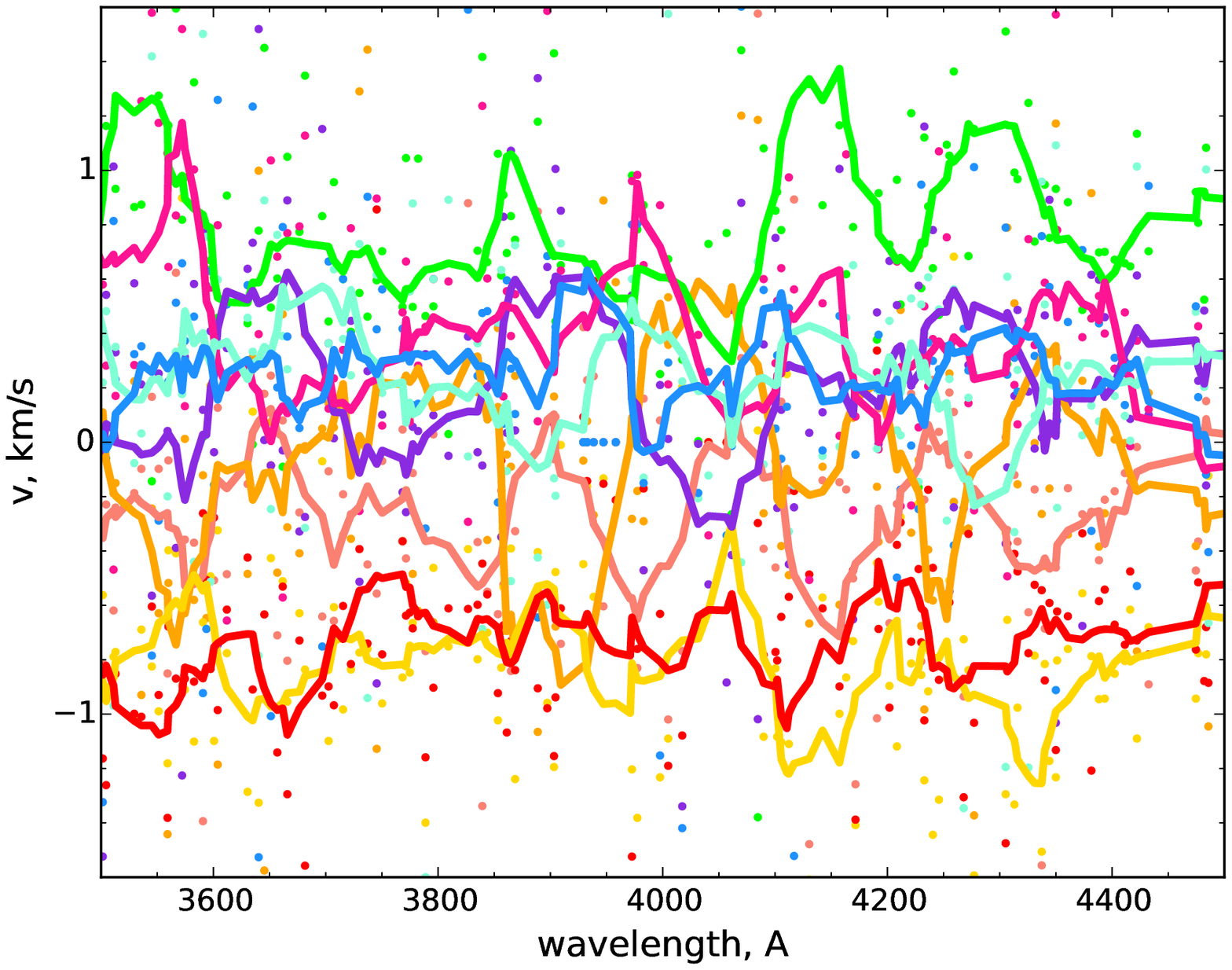} \\
	\end{tabular}
	\caption{{\sl Left}: The histogram of the measured shifts for exposures number 3, 4 and 5 (see Table~\ref{tab:obs}) in the cross calibration procedure. {\sl Right}: The dependence of the measured shifts from wavelength for each of used exposures (shown by different colors). The points show measured values of relative shifts, while the lines with corresponding colors indicate smoothed interpolation of the data over several adjacent points. For illustration purposes, the truncated region of spectrum is shown, since absorption lines in this region are dense enough.}
	\label{fig:fig1}
\end{figure*}

It is well established that echelle spectra obtained by HIRES/KECK or UVES/VLT and calibrated with a ThAr lamp show both velocity offsets between different exposures and intra-order velocity distortions within each exposure \citep[see, e.\,g.,][]{Griest2010, Whitmore2010, Wendt2011, Molaro2013, Rahmani2013, Evans2014}. The main reasons of these calibration
distortions are an unstable position of the source in the slit during exposure, difference in position between the exposure and the calibration lamp, changes in conditions between an exposure and a calibration procedure, non-linearity of echelle orders, and non-uniformity of ThAr lines which are used for the reference. To take into account improper wavelength calibration, a procedure similar to one described by \cite{Evans2013} has been used as follows. First of all, bad pixels (mainly related to cosmic rays) in each exposure have been removed using iterative smoothing of the spectra and visual inspection. After that, the exposures have been coscaled in each pixel using comparison of their mean fluxes calculated in a relatively large (5\,\AA) sliding windows. Then, all the exposures have been convolved with the Gaussian function with a constant full width at half maximum in velocity space corresponding to the resolution of each exposure. The convolved spectra have been interpolated by a polynomial using Neville's algorithm to conserve the local flux in the bin \citep[see][]{Wendt2011}. Finally, the exposures have been compared with each other using $\chi^2$ criteria \citep[see][]{Evans2013} in predetermined spectral regions. The latter were defined using regions of absorbed flux (mostly of Ly\,$\alpha$ forest absorption lines) and separated by unabsorbed parts of spectra. The results of the wavelength recalibration are shown in Fig.~\ref{fig:fig1}. The exposures have been found to be shifted up to 3 km/s relatively to each other (see the Right panel of Fig.~\ref{fig:fig1}). In addition, there is up to 1 km/s dispersion of the shifts within the each exposures (see the Left panel of Fig.~\ref{fig:fig1}). These results are in the quantitative agreement with a result of \cite{Griest2010} who found similar shifts for HIRES/KECK. All the exposures have been corrected applying the measured shifts in each determined spectral region before coadding the spectrum.

\setlength{\tabcolsep}{0pt}
\setlength{\tabcolsep}{3pt}

\section{Analysis}

\subsection{Fitting procedure}

Relevant absorption lines have been fitted with the standard multi-component Voigt profile procedure. To obtain the best fit and errors on the derived parameters, a software package written by SB was used. This code uses the $\chi^2$ maximum likelihood function to compare a fit model with the observed spectrum. The best fit model and statistical errors are obtained by the Markov Chain Monte Carlo (MCMC) technique which uses an affine invariant sampler \citep{Goodman2000}. A similar approach was already used for absorption line analysis \citep{King2009}. This method allows the parameter space to be explored accurately and the global maximum of a likelihood function to be found even for high dimensions of parameter space. This sufficiently improves  Levenberg-Marquardt least square minimization which tends to stop at local minimums with increasing parameter space. An additional advantage of the method is that it allows one to derive a shape of the likelihood function, which in some cases is asymmetric. This results in asymmetric errors for the derived parameters. In what follows, best fit parameters and their errors correspond to the maximum of the likelihood function and the 68.3\% quantile interval near the maximum, respectively. The latter is formally 1$\sigma$ interval for normal distribution.

\subsection{SubDLA system at $z=2.437$}
\label{subs:subDLA}

The subDLA (sub Damped Lyman Alpha system -- an absorption system with H\,{\sc i} column density between $10^{19}$ and $10^{20.3}\,$cm$^{−2}$, \citealt{Dessauges-Zavadsky2003,Peroux2003}) at $z_{\rm abs}=2.437$ was detected in this spectrum by \cite{Carballo1995}. Fitting the H\,{\sc i} Ly\,$\alpha$ lines with a simple one-component model gives the total H\,{\sc i} column density of log\,$N$(H\,{\sc i}$)=19.983\pm0.010$ with redshift of the component $z=2.437225(7)$. Since the HIRES spectrum is not flux calibrated and unabsorbed quasar continuum is not properly known, the continuum in the region of Ly\,$\alpha$ lines has been fitted simultaneously with the line profile using the 15-order Chebyshev polynomial. Due to high saturation of \ion{H}{I} lines, the component structure of the absorption system cannot be defined via the H\,{\sc i} absorption line analysis. The component structure of the system can be defined by studying associated absorption lines of metals. There is a plenty of metals which show absorption lines associated with this subDLA, namely, O\,{\sc i}, Si\,{\sc ii}, C\,{\sc ii}, Al\,{\sc ii}, Fe\,{\sc ii}, C\,{\sc iv}, Si\,{\sc iv} and others. The next section gives an overview of the velocity component structure of the subDLA derived from metal line profiles.

\begin{figure}
\includegraphics[width=\linewidth]{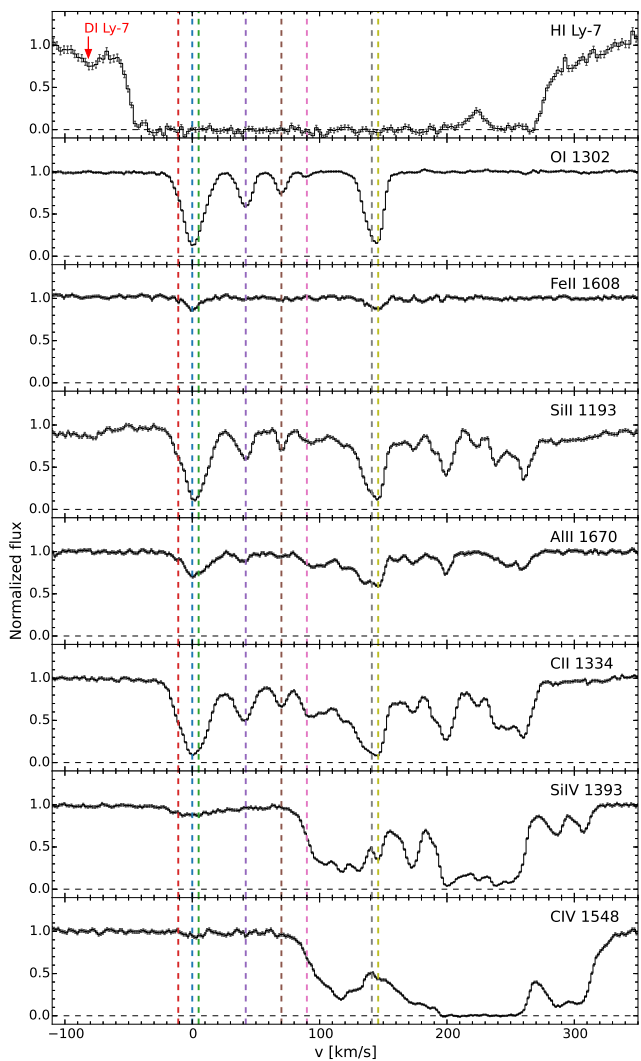}
\caption{Line profiles of H\,{\sc i} Ly-7 and metal species with various ionization potentials associated with the subDLA at $z_{\rm abs}=2.437$. Dashed vertical lines show the dominant components in the velocity structure of \ion{O}{I} lines associated with the subDLA. The x-axis is the velocity offset from the main component (indicated by a blue dashed vertical line) of the subDLA.
}
\label{fig:fig2}
\end{figure}

\subsection{Component structure of the subDLA}

This subDLA exhibits a relatively reach velocity structure, which sufficiently complicates the estimation of the D/H ratio. Line profiles of species with different values of ionization potential are shown in Fig.~\ref{fig:fig2}. The component structure depends on the ionization potential of the species -- ones with similar ionization potentials show a similar velocity structure. This situation is typical for subDLAs and DLAs. To achieve our main goal, estimation of the D/H ratio, we will focus on the species with low values of the ionization potential: O\,{\sc i}, Si\,{\sc ii} and C\,{\sc ii}. It is reasonable, since these species correspond to a neutral fraction of the interstellar medium, i.\,e. the major fraction of neutral hydrogen and deuterium is associated with these species. Moreover, the absorption lines from these species show the dominant component, in which the D/H ratio can be constrained.

There are at least eight components in the O\,{\sc i} profile (see Fig.~\ref{fig:fig2} and Section~\ref{subs:model}) and at least seventeen ones in Si\,{\sc ii}. It is not surprising that O\,{\sc i}, Si\,{\sc ii}, C\,{\sc ii} and Fe\,{\sc ii} have different numbers of components, since their ionization potentials differ from each other. 

Neutral hydrogen column density is high enough in components of the subDLA, therefore all the H\,{\sc i} Ly-series lines are blended with each other. Lyman series of D\,{\sc i} lines are shifted relatively to the corresponding H\,{\sc i} lines by about $-81.6$\,km/s. As a result, the D\,{\sc i} line profiles are blended by H\,{\sc i} for Ly\,$\alpha$ and Ly\,$\beta$ lines, but seen in the blue wing of higher Ly-series H\,{\sc i} lines. The D\,{\sc i} profiles corresponding to the three bluer components are clearly identified in the Ly\,$\gamma$, Ly\,$\delta$, Ly-5, Ly-7, Ly-8, Ly-9 lines. These lines have been used to estimate D\,{\sc i} column density. To estimate H\,{\sc i} column density and consequently the D/H ratio, a model presented in the next section has been used.


\subsection{Model}
\label{subs:model}

Deuterium abundance can be measured in the three bluer components only, since in other components the D\,{\sc i} lines are blended by corresponding H\,{\sc i} lines. Six D\,{\sc i} lines have been used to estimate the D\,{\sc i} column density: Ly\,$\gamma$, Ly\,$\delta$, Ly-5, Ly-7, Ly-8 and Ly-9. For better constraint of the Doppler parameter, $b$, of the D\,{\sc i} lines, the lines are fitted simultaneously with corresponding H\,{\sc i} and low-ionization metal lines of O\,{\sc i} (1302\,{\AA} and 1039\,\AA), Si\,{\sc ii} (1526\,\AA, 1304\,\AA, 1260\,\AA, 1193\,\AA, 1190\,\AA\ and 989\,\AA) and C\,{\sc ii} (1334\,\AA\ and 1036\,\AA). For each (of three) fitted bluer component, a turbulent $b$-parameter, $b_{\rm turb}$, and temperature of the gas, $T$, have been varied. Therefore, the Doppler parameter of each species with atomic weight, $M$, can be calculated as
\begin{equation}
\label{doppler}
b=\sqrt{b_{\rm turb}^2+2T/M}.
\end{equation}

\begin{figure*}
\includegraphics[width=1.0\linewidth]{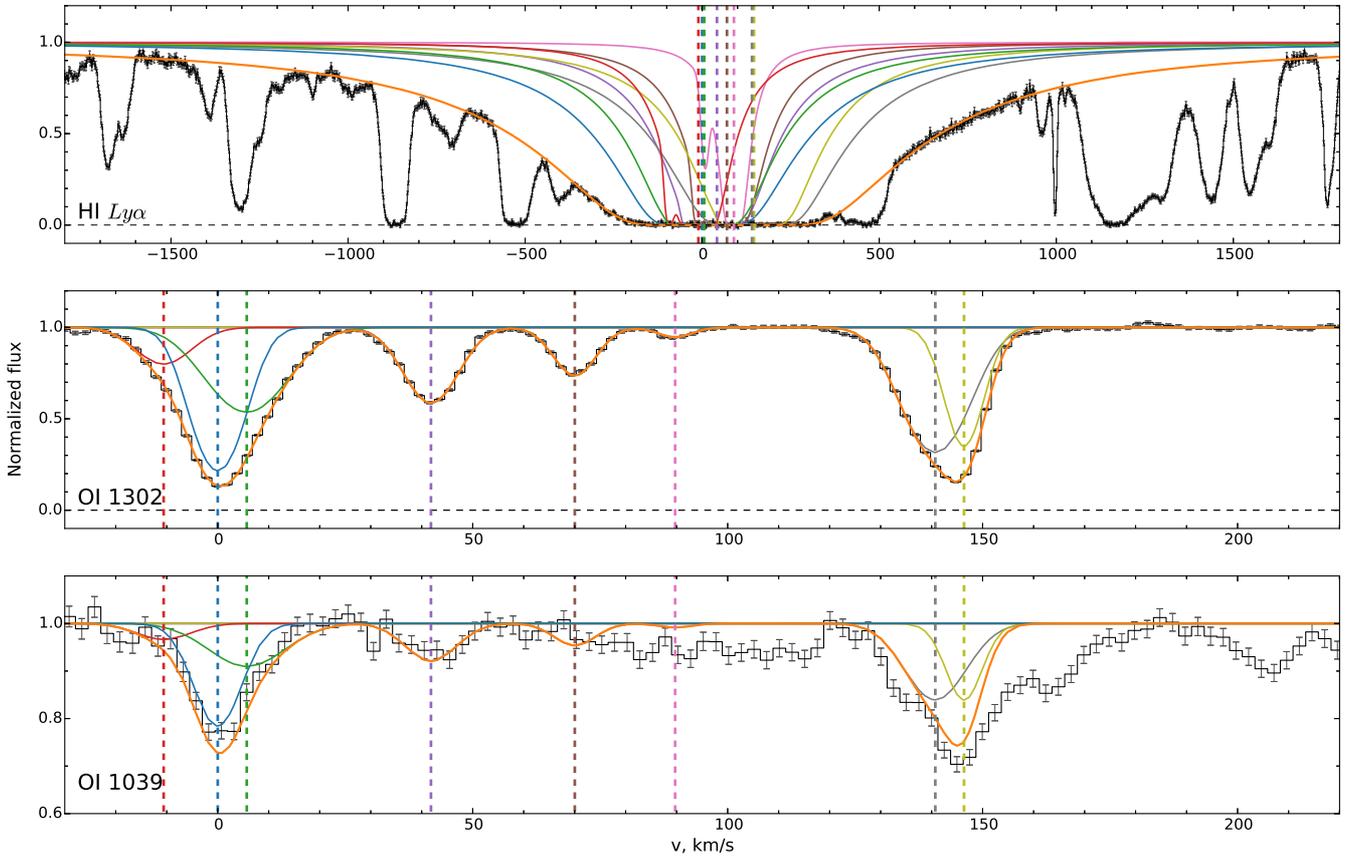}
\caption{Line profiles of Ly\,$\alpha$, O\,{\sc i} 1302 and O\,{\sc i} 1039 lines. The black line shows the spectrum of the QSO J\,1444$+$2919. The orange line shows the total profile of the best fit model, while the other lines represent profiles of each component separately. The profile of Ly\,$\alpha$ lines is already corrected for the continuum which was reconstructed during the fit (see text).}
\label{fig:fig3}
\end{figure*}

\begin{figure}
\includegraphics[width=\linewidth]{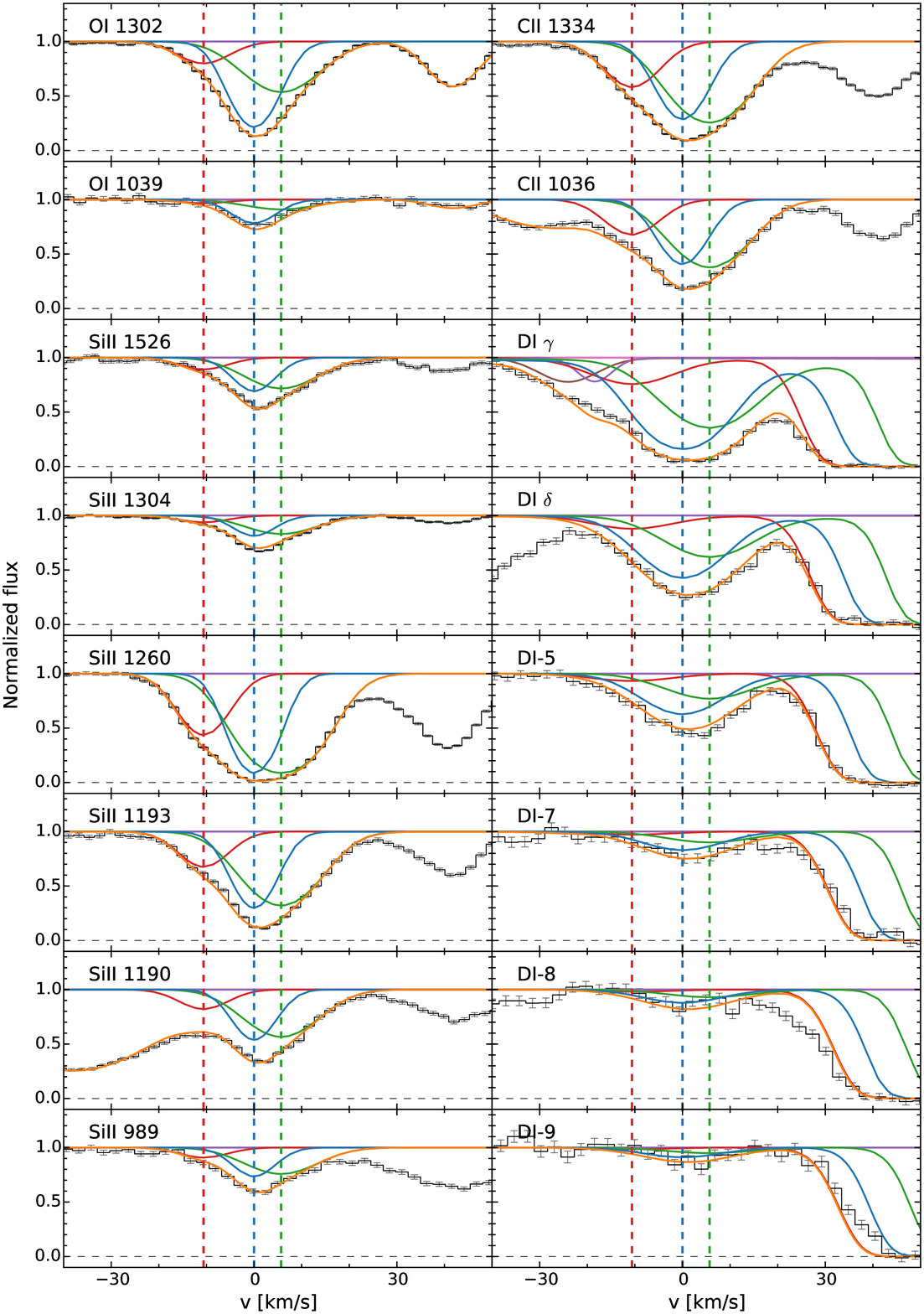}
\caption{Line profiles of the absorption lines (indicated in the left-top corner of each panel) used in the analysis. The colors of the lines are the same as in Fig.~\ref{fig:fig3}.}
\label{fig:fig4}
\end{figure}

Since the subDLA has a complex velocity component structure, the associated H\,{\sc i} column densities for the three bluer components cannot be derived from fitting the Ly-series lines of the subDLA without any additional assumptions. To estimate H\,{\sc i} column densities of the three bluer components, O\,{\sc i} lines have been used, since the O\,{\sc i} ionization potential of 13.618\,eV is very close to that of hydrogen, 13.589\,eV, and therefore these species are expected to be co-spatial. To this end, 5 additional components seen in O\,{\sc i} line profile have been added to the fit model and O/H ratio has been fixed to be the same for all the components. In other words, metallicity of oxygen is assumed to be the same over all the subDLA components. This is a rough assumption, but it is usually used in similar estimations \citep[see, e.\,g.,][]{Noterdaeme2012}. We estimate systematic error introduced by this assumption in the derived D/H ratio in Section~\ref{systematics}. With this assumption, H\,{\sc i} Ly\,$\alpha$ lines have been added to constrain H\,{\sc i} column density. Since the spectrum is not properly flux-calibrated, the continuum in the region of H\,{\sc i} Ly\,$\alpha$ has been varied using the 15-order Chebyshev polynomial (as was discussed in Section~\ref{subs:subDLA}). The line profiles of the best fit are shown in Fig.~\ref{fig:fig3} and Fig.~\ref{fig:fig4}. The values of the best fit parameters and their errors are given in Table~\ref{tab:fit_results}. The value $\chi^2/{\rm dof} = 4932/3045\approx1.6$ for the best fit model is rather good, but we suggest that there are some systematic errors (e.\,g. unaccounted blends and/or continuum variations) which are not taken into account. A naive expectation is that these systematic errors lead to increase of the obtained errors by a factor of $\sqrt{1.6}\approx1.26$.

\begin{figure}
\includegraphics[width=\linewidth]{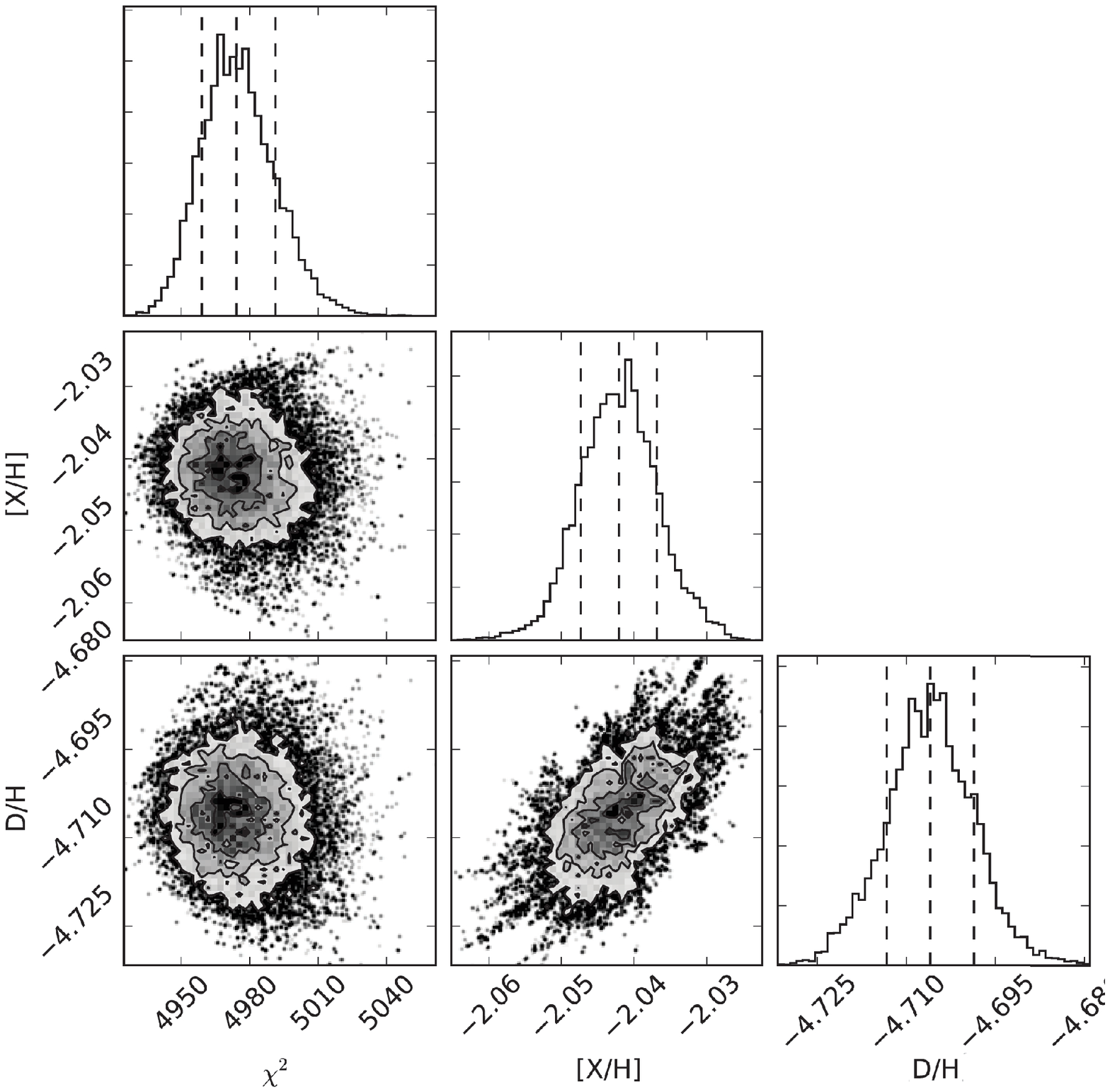}
\caption{Triangle plot of the 2D posterior distribution. For illustrative purposes, only two main parameters (out of 50 that were varied) of the fit model are shown: overall metallicity of oxygen in the subDLA, [O/H], and the D/H ratio.
The diagonal panels show 1D posterior distribution for the specified parameters.}
\label{fig:fig5}
\end{figure}

Fig.\ref{fig:fig5} shows the 2D posterior distribution of likelihood function for metallicity and D/H ratio derived in our model. The metallicity (via oxygen) of the subDLA has been found to be [O/H]\,=\,$-2.042\pm0.005$. It implies that the astration correction for D is negligible. A close value of the metallicity can be obtained from the determined total O\,{\sc i} column density of all 8 components, log\,$N$(O\,{\sc i}$) = 14.526^{+0.066}_{-0.023}$, and the total H\,{\sc i} column density obtained from a one-component fit. This suggests that a simple estimate of H\,{\sc i} column densities for the blended components can be made using ratios of the measured O\,{\sc i} column densities in the components. The D/H ratio has been found to be
\begin{equation}
{\rm log\,(D/H)}=-4.706\pm0.007_{\rm stat}.
\end{equation}
This estimate includes statistical errors only which are comparable (or even less) with recent, the most precise, D/H measurements of \citet{Pettini2012, Cooke2014}. The obtained result does not include any systematic errors concerning a choice of the appropriate fit model. The smallness of the statistical errors is a result of the very high S/N ratio (up to $\sim 100$) in the studied spectrum. 

\begin{table*}
	\begin{center}
		\begin{tabular}{ccccccccc}
			\hline
			No. & z & v$^\dag$ & $b_{\rm turb}$, km/s & $T$, K & $\log{N} ($\ion{O}{I}) & $b_ {\rm \ion{O}{I}}$ $^\ddag$  , km/s & $\log{N} ($\ion{Si}{II}) & $\log{N} ($\ion{C}{II}) \\
			\hline
			1  &  2.4364481(4)  &  -11.3  &  $ 5.38^{+0.39}_{-0.26}$  &  $11672^{+  174}_{-  207}$  &  $13.143^{+0.054}_{-0.040}$  &    &  $12.363^{+0.046}_{-0.039}$  &  $13.155^{+0.039}_{-0.035}$  \\
			2  &  2.4365772(1)  &    0.0  &  $ 4.74^{+0.19}_{-0.20}$  &  $ 9861^{+  633}_{-  637}$  &  $14.044^{+0.024}_{-0.018}$  &    &  $12.874^{+0.049}_{-0.039}$  &  $13.524^{+0.052}_{-0.043}$  \\
			3  &  2.4366362(7)  &    5.1  &  $10.01^{+0.25}_{-0.37}$  &  $ 8466^{+ 1486}_{- 1210}$  &  $13.869^{+0.031}_{-0.046}$  &    &  $13.081^{+0.027}_{-0.042}$  &  $13.773^{+0.028}_{-0.042}$  \\
			4  &  2.4370573(1)  &   41.9  &    &    &  $13.616^{+0.005}_{-0.004}$  &  $ 6.31^{+0.10}_{-0.09}$  &    &    \\
			5  &  2.4373811(1)  &   70.1  &    &    &  $13.291^{+0.008}_{-0.009}$  &  $ 4.94^{+0.17}_{-0.14}$  &    &    \\
			6  &  2.4376082(5)  &   89.9  &    &    &  $12.540^{+0.034}_{-0.032}$  &  $ 4.28^{+0.79}_{-0.81}$  &    &    \\
			7  &  2.4381935(3)  &  141.0  &    &    &  $14.023^{+0.021}_{-0.025}$  &  $ 7.76^{+0.12}_{-0.14}$  &    &    \\
			8  &  2.4382528(1)  &  146.2  &    &    &  $13.770^{+0.038}_{-0.036}$  &  $ 3.60^{+0.25}_{-0.29}$  &    &    \\
			\hline
			$\rm [O/H]$ & $-2.042\pm0.005$ & \multicolumn{6}{c}{} \\
			$\rm D/H$ & $-4.706\pm0.007$ & \multicolumn{6}{c}{} \\
			\hline
		\end{tabular}
		\caption{Fitting results.
			$^\dag$ Relative to the second component at  z=2.4365772. $^\ddag$ Doppler parameters of \ion{O}{I} for components 1, 2 and 3 are constrained by $b_{\rm turb}$ and $T$ using Eq.~\eqref{doppler}.}
		\label{tab:fit_results}
	\end{center}
\end{table*}

\subsection{Systematic errors}
\label{systematics}

There are several possible sources of systematic errors such as contaminations by blends, unaccounted components in the velocity structure, continuum placement uncertainties (see Section~\ref{subs:errors}). Of these, the major source of systematic error for the D/H value derived in line profile analysis of the studied subDLA comes from an assumption used during the fitting procedure that all the components in the subDLA have equal metallicity, [O/H]. Such an assumption is rough, since the components of the subDLA have sufficient velocity shifts ($\sim150$ km/s) and therefore are likely not spatially connected, i.\,e. correspond to spatially remote regions. Therefore it is reasonable to expect difference in the metallicity. It is well known that disks of some galaxies show variation of [O/H] with 0.2 dex and higher (see, e.\,g., \citealt{vanZee1998, Moustakas2010}). Hence, our assumption could introduce bias in the derived D/H value. We assessed the corresponding systematic error as follows.

\begin{figure*}
	\includegraphics[width=1.0\linewidth]{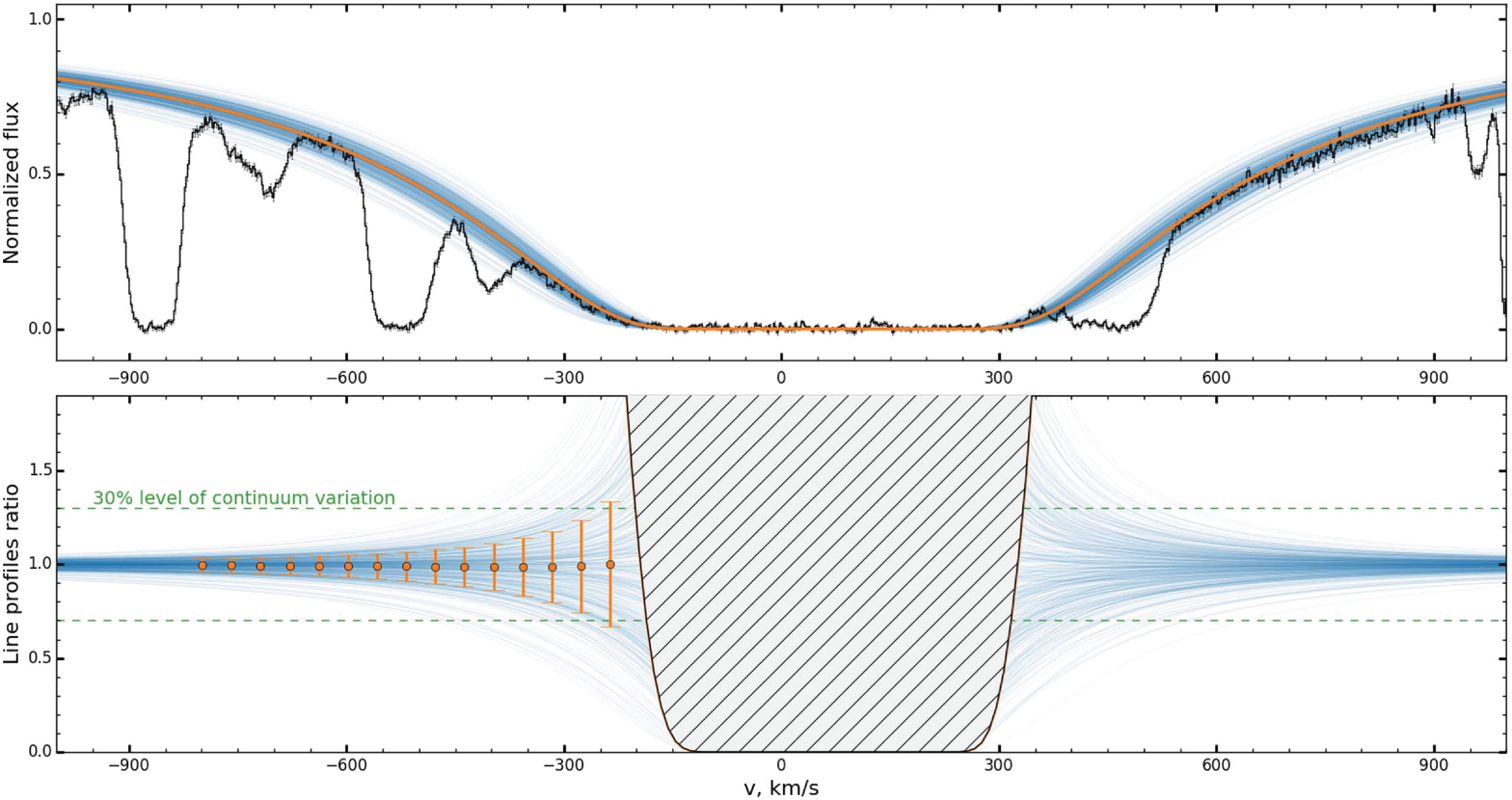}
	\caption{{\sl Top panel:} The line profile of Ly\,$\alpha$ line. The orange line corresponds to the best fit model with uniform [O/H] metallicity over the components of the subDLA system. The blue lines show the variation of profile of Ly\,$\alpha$ in case of metallicity in the components randomly varies at 0.1 dex level relatively to the best fit value. {\sl Bottom panel:} The blue lines show the ratio of the disturbed line profiles to the best fit model line profile. The dependence of mean and standard deviation of this ratio versus position of line profile are shown by orange points and error bars, respectively. The hatched region corresponds to the region where it is not reasonable to compare line profiles, since in this region values of the flux are less than two average errors of the flux at certain wavelength.}
	\label{systematic}
\end{figure*}

The column densities of \ion{O}{I} in each component are accurately measured, since line profiles of \ion{O}{I} are not saturated. The main difficulties in estimating such systematic error are that components of Ly\,$\alpha$ line are blended with each other and that we do not know exact continuum in the region of Ly\,$\alpha$ line (continuum shape was an independent variable during the fit, i.\,e. continuum was reconstructed). In principle, it can be done by means of Monte-Carlo simulations applying random variations in the metallicity, [O/H], of the components and finding the best fit for each realization. However, such simulations are very time consuming. Therefore, to estimate the systematic error corresponding to metallicity variations in the components of the subDLA system we used the following procedure. We examine how the profile of \ion{H}{I} Ly\,$\alpha$ line changes if the metallicity, [O/H], is allowed to vary over the subDLA components. We make runs, each of them consists of many realizations, where we apply random normally distributed variations in the metallicity of each subDLA component with zero mean and standard deviation at some specified level for the run. One of this run is shown in the upper panel of Fig.~\ref{systematic}. For each of the run we estimate how the profile of Ly\,$\alpha$ line changes relatively to the line profile of the best fit model (with constant metallicity over components) and calculate the dependence of the mean and dispersion of their ratio on the wavelength (see the lower panel of Fig.~\ref{systematic}). It was found that dispersion of this ratio gradually increases with approaching to the bottom of the line and with increase of the specified level of the metallicity variations at the run. If we try to find the best fit model for some particular realization then, roughly speaking, variations of the line profiles shown in the bottom panel of Fig.~\ref{systematic} will translate mostly to the variations of the continuum shape at the same value. Therefore, we can constrain the possible variation of the metallicity assuming that continuum variations cannot be higher than some level. We assume that 20\% continuum variation on such short wavelength range (less than 3\AA) is very unlikely, even for the studied echelle spectrum which is not properly flux calibrated. However, we use 30\% as an upper limit to obtain a conservative value of systematic error. This corresponds to the run where standard deviation of metallicity variation is 0.10 dex and shown in the bottom panel of Fig.~\ref{systematic}. This transforms to 0.067 dex variation of the total \ion{H}{I} column density for three components (1, 2 and 3) where D/H ratio was constrained. We conclude that the D/H systematic error related to the possible [O/H] variations in the components of the subDLA system is 0.067 dex. Surely, this estimate is somewhat arbitrary, since it depends on specified level of possible continuum variations, but it gives typical values.

As a result, the obtained D/H value which includes both systematic and statistical errors is
\begin{equation}
{\rm log\,(D/H)}=-4.706\pm0.007_{\rm stat}\pm0.067_{\rm syst}
\end{equation}
or
\begin{equation}
{\rm D/H}= \left(1.97^{+0.33}_{-0.28}\right) \times 10^{-5}.
\end{equation}

\section{The primordial deuterium problem}
\label{sec:D_problem}

\begin{table*}
 \centering
 \begin{minipage}{140mm}
  \caption{Summary of D/H values based on D\,{\sc i}/H\,{\sc i} lines measured in QSO absorption systems. Here $z_{\rm em}$ and $z_{\rm abs}$ are redshifts of the QSO and the absorption system respectively; [X/H] is metallicity with respect to solar and X is a used element; log$\,N$(H\,{\sc i}) is logarithm of total H\,{\sc i} column density of the absorption system, measured in cm$^{-2}$; D/H is relative deuterium abundance in the absorption system.}
  \label{tab:D_H_data_sample}
   \begin{tabular}{@{}lllcclllr@{}}
  \hline
 Quasar & $z_{\rm em}$ & $z_{\rm abs}$ & [X/H] & X & log$\,N$(H\,{\sc i}) & D/H ($\times 10^{-5}$) & References$^{a}$ \\
 \hline
 Q\,0105$+$1619   & 2.64 & 2.536 & -1.77   & O      & $19.426\pm 0.006$  & $2.58^{\,+\,0.16}_{\,-\,0.15}$ & \citet{Cooke2014} \citep{OMeara2001} \parbox[c][4.5mm][c]{0mm}{}\\
 Q\,0347$-$3819   & 3.23 & 3.025 & 0.98$^b$& Zn$^b$ & $20.626\pm 0.005$  & $3.75\pm 0.25$ & \citet{Levshakov2002} \parbox[c][4.5mm][c]{0mm}{}  \\
 \,J\,0407$-$4410 & 3.02 & 2.621 & -1.99   & O      & $20.45\pm 0.10^{e}$& $2.8^{\,+\,0.8}_{\,-\,0.6}$ &  \citet{Noterdaeme2012} \parbox[c][4.5mm][c]{0mm}{} \\
 Q\,0913$+$0715   & 2.79 & 2.618 & -2.40   & O		& $20.312\pm 0.008$  & $2.53^{\,+\,0.11}_{\,-\,0.10}$ & \citet{Cooke2014} \citep{Pettini2008b} \parbox[c][4.5mm][c]{0mm}{} \\
 Q\,1009$+$2956   & 2.63 & 2.504 & -2.5    & Si 	& $17.39\pm 0.06$ 	 & $4.0^{+0.6}_{-0.7}$ & \citet{Burles1998b} \parbox[c][4.5mm][c]{0mm}{} \\
 \,J\,1134$+$5742 & 3.52 & 3.411 & $<$-4.2 & Si 	& $17.95\pm 0.05$    & $2.0^{\,+\,0.7}_{\,-\,0.5}$ & \citet{Fumagalli2011} \parbox[c][4.5mm][c]{0mm}{} \\
 Q\,1243$+$3047   & 2.56 & 2.526 & -2.79   & O		& $19.73\pm 0.04$	 & $2.42^{\,+\,0.35}_{\,-\,0.25}$ & \citet{Kirkman2003} \parbox[c][4.5mm][c]{0mm}{} \\
 \,J\,1337$+$3152 & 3.17 & 3.168 & -2.68   & Si 	& $20.41\pm 0.15$	 & $1.2^{\,+\,0.5}_{\,-\,0.3}$ & \citet{Srianand2010} \parbox[c][4.5mm][c]{0mm}{} \\
 \,J\,1358$+$6522 & 3.17 & 3.067 & -2.33   & O  	& $20.495\pm 0.008$  & $2.58\pm 0.07$ & \citet{Cooke2014}    \parbox[c][4.5mm][c]{0mm}{} \\
 \,J\,1419$+$0829 & 3.03 & 3.050 & -1.92   & O  	& $20.392\pm 0.003$  & $2.51\pm 0.05$ & \citet{Cooke2014} \citep{Pettini2012} \parbox[c][4.5mm][c]{0mm}{} \\
 \,J\,1444$+$2919 & 2.66 & 2.437 & -2.04   & O      & $19.983\pm 0.010$  & $1.97^{\,+\,0.33}_{\,-\,0.28}$ & This work \parbox[c][4.5mm][c]{0mm}{} \\
 \,J\,1558$-$0031 & 2.82 & 2.702 & -1.55   & O  	& $20.75\pm 0.03$    & $2.40^{\,+\,0.15}_{\,-\,0.14}$ & \citet{Cooke2014} \citep{OMeara2006} \parbox[c][4.5mm][c]{0mm}{} \\
\multirow{2}{*}{Q\,1937$-$1009$^c$} & \multirow{2}{*}{3.79} & 3.256 & -1.87   & O  & $18.09\pm 0.03$ & $2.45^{\,+\,0.30}_{\,-\,0.27}$           & \citet{RiemerSorensen2015} \citep{Crighton2004} \parbox[c][4.5mm][c]{0mm}{} \\
                 &      & 3.572  & $<$-0.9 & O  & $17.86\pm 0.02$ & $3.3\pm 0.3$ & \citet{Burles1998a} \parbox[c][4.5mm][c]{0mm}{} \\
 Q\,2206$-$199   & 2.56 & 2.076  & -2.04$^d$& O$^d$ & $20.436\pm 0.008$  & $1.65\pm 0.35$ & \citet{Pettini2001} \parbox[c][4.5mm][c]{0mm}{} \\
\hline 
\end{tabular}

\small{$^a$Works in parentheses correspond to previous measurements of the D/H value for the same absorption systems.}
\small{$^b$Measured by \citet{Ledoux2003}.}
\small{$^c$There are two absorption systems at z$_{\rm abs}$=3.256 and z$_{\rm abs}$=3.572 towards J\,1939$-$100 with identified D\,{\sc i} lines.}
\small{$^d$Measured by \citet{Pettini2008a}.}
\end{minipage}
\end{table*}

\begin{figure}
\includegraphics[width=84mm]{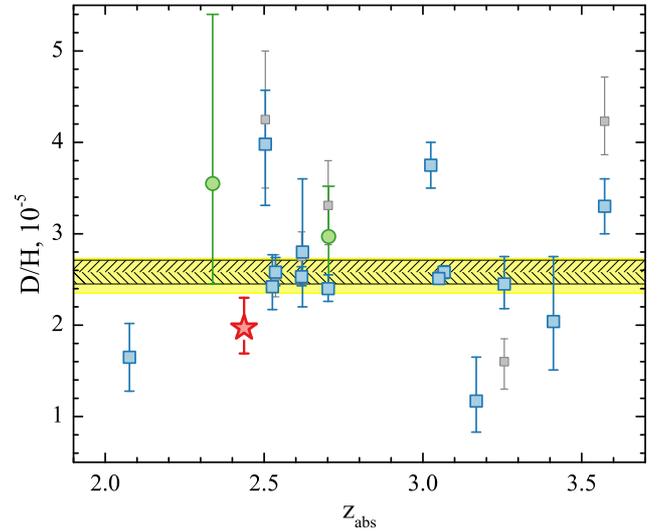}
\caption{D/H values measured in QSO absorption line systems as a function of absorption redshift. The red star represents the new case reported here (J\,1444$+$2919). 
Squares are the measurements based on D\,{\sc i}/H\,{\sc i} lines taken from the literature: the blue ones correspond to the values used in calculation of the mean value and presented in Table~\ref{tab:D_H_data_sample}, while the grey ones are the earlier estimates. 
The green circles are the HD/2H$_2$ based measurements in two DLAs presented in the literature. The bands mark the corresponding unweighted mean value (yellow) and the standard BBN prediction based on the Planck data (hatched) with their 68.3\,\% confidence intervals.}
\label{fig:D_H_ratio}
\end{figure}

For the moment, there are 15 (including this work) measurements of the primordial D/H value. The main data about the absorption systems which the primordial deuterium abundance was determined for are summarised in Table~\ref{tab:D_H_data_sample}.
Fig.~\ref{fig:D_H_ratio} presents all the available measurements of the primordial deuterium abundance listed in Table~\ref{tab:D_H_data_sample}. For illustration, the D/H values based on the HD/2H$_2$ measurement technique \citep{Balashev2010, Ivanchik2010} are also presented. A conspicuous feature of the sample is that scatter of the D/H values about their mean value considerably exceeds the errors of the individual measurements what has been already mentioned by many authors \citep[e.\,g.,][]{Kirkman2003, Pettini2008b, Ivanchik2010, Olive2012}. The most plausible reason for this is that errors in some (or even in all) measurements have been underestimated.

\subsection{Sources of uncertainties}
\label{subs:errors}

A challenge of deciphering the structure of the absorption system and its physical parameters encounters several sources of uncertainties. Their extensive, but by no means complete, list was discussed by \cite{Kirkman2003}. Of these, there are several ones which affect the obtained D/H value to a greater extent.

First of all, by adopting a different number of velocity components in the absorption system, the obtained D/H values can differ significantly. \cite{RiemerSorensen2015} demonstrated recently how a high S/N ratio spectrum can help to overcome this obstacle. Their work along with their previous measurement of the primordial deuterium abundance for the same absorption system but based on a low S/N spectrum \citep{Crighton2004} illustrates the strong dependence of the obtained D/H value on the adopted multicomponent structure. In this paper we have studied an assumption of constant metallicity over components of an absorption system as a source of addition uncertainties. Using this assumption during the fitting procedure draws a possible error in the deduced D/H value especially for systems with complex velocity structure. Another source of uncertainties roots in the adopting model for the velocity distribution of the absorbing gas. Using micro- or mesoturbulent models results in D/H values which differ significantly from each other as well, what has been shown by \cite{Levshakov1998, Levshakov2000}. Finally, addition classes of systematic errors stem from the effect of continuum placement uncertainty on the H\,{\sc i} column density and from contamination by blends with some Ly\,$\alpha$ forest lines \citep[see, e.\,g.,][]{Kirkman2003}.

A difference between D/H values stemming from different adopted models illustrates the real systematic uncertainties, which are, however, difficult to be estimated properly. High S/N ratio spectra may be crucial for narrowing the available parameter space and, thus, allow a correct model to be chosen \citep[e.\,g.,][]{RiemerSorensen2015}. We note that the spectrum studied in this paper has very high S/N ratio.

Note that there is another viable explanation of the observed scatter of the D/H values. It is possible that the dispersion is real. However, no reason for this is known for the moment \citep[see., e.\,g.,][]{Klimenko2012}. As metallicity is low in these systems (see Table~\ref{tab:D_H_data_sample}), an astration correction for D is assumed to be negligible. In either case, more data are needed to find it out. In case of realness of the observed scatter, the primordial deuterium abundance can be found as a plateau in the D/H measurements \citep[e.\,g.,][]{RiemerSorensen2015}. The latter is somewhat we have now in case of the primordial lithium abundance \citep[see, e.\,g.,][]{Sbordone2010}.

\subsection{The Precision Sample}

A set of selection criteria was suggested recently by \citet{Cooke2014} aimed at identification of few number of the absorption systems where the most accurate and precise measurements of the primordial D/H value are potentially possible. In addition, reanalysis of four systems and an analysis of a new damped Ly$\,\alpha$ (DLA) system which satisfy these criteria were done as well as an unprecedentedly precise value of the primordial deuterium abundance was determined \citep{Cooke2014}. In addition, the power of high-precision measurements of the primordial deuterium abundance in testing extensions of the Standard Model was demonstrated.

The new system reported here does meet these criteria. However, the obtained D/H value agrees with the weighted mean D/H value for the {\it Precision Sample} obtained by \cite{Cooke2014} at $2\sigma$ level only. This can be a random fluctuation or could be a consequence of that the errors in some D/H measurements of \citet{Cooke2014} could be underestimated. The latter has already been mentioned by \citet{RiemerSorensen2015}.
	
A new {\it Precision Sample}, consisting of that of \cite{Cooke2014} and of the new result reported here, leads to a weighted mean value of $\langle{\rm D/H}\rangle = (2.52\pm0.04)\times10^{-5}$ (the larger of the errors has been adopted for the calculation of weighting in cases of asymmetric errors on the individual D/H measurements). 
Given that the systematic error of the new D/H value reported here is significantly higher than that of the measurements reported by \citet{Cooke2014}, it does not vary the weighted mean D/H value for the {\it Precision Sample} sizably.

Whilst the approach of \cite{Cooke2014} does provide a precise measurement, reliance on only a few systems is risky as one can be misled to a wrong plateau in the deuterium abundance values. Possible underestimation of systematic errors can bring bias into the mean value (especially if use a weighted mean). Hence, it might be reasonable to relax the set of strict selection criteria and, thus, increase the number of acceptable absorption systems with measured D/H values. It might be vital to minimize a possible effect of underestimation of systematics as well as to understand the real reasons of the observed scatter in the D/H data sample, what have also been stated by \citet{RiemerSorensen2015}.

\subsection{The primordial deuterium abundance}

For the reasons described in the previous section, the whole D/H data sample of 15 listed in Table~\ref{tab:D_H_data_sample} is used to calculate the primordial deuterium abundance. For systems, which several measurements of the deuterium abundance have been done for, the D/H values are treated as follows. Four remeasured D/H values obtained by \citet{Cooke2014} are used as they are more precise than the previous estimates \citep{OMeara2001, OMeara2006, Pettini2008b, Pettini2012}. Despite the possibility of underestimation of the continuum placement uncertainty for the $z_{\rm abs}=3.050$ DLA towards QSO~J\,1419$+$0829 \citep{Cooke2014} mentioned by \cite{RiemerSorensen2015}, there is no objective reason to ignore this measurement. The revised D/H value at the $z_{\rm abs}=3.256$ absorption system towards Q\,1937$-$1009 reported recently by \citet{RiemerSorensen2015} is used as in their analysis a spectrum with a higher S/N ratio was treated and, as a result, more components of the absorption system were detected than in their previous analysis \citep{Crighton2004}.

Given that the uncertainties in some measurements could be underestimated, it is likely to bring a systematic bias if the primordial deuterium abundance is determined via a weighted mean value. Alternatively, an unweighted mean D/H value might be more appropriate for this purpose.

An unweighted mean value of the D/H ratio over the whole data sample (15 measurements) gives a conservative estimate of the primordial deuterium abundance
\begin{equation}
({\rm D/H})_{\rm p}=(2.54\pm 0.19)\times10^{-5},
\end{equation}
which is in good agreement with the primordial value of (D/H)$_{\rm BBN}=(2.58\pm 0.13)\times10^{-5}$ \citep{Cyburt2015} predicted by standard BBN using $\eta$ derived independently from an analysis of the cosmic microwave background (CMB) anisotropy \citep{Planck2015}. Note that the unweighted mean of 15 D/H measurements is in excellent agreement with the weighted mean based on the new {\it Precision Sample}. At the same time, a confident interval of the former is less stringent than that obtained for the new {\it Precision Sample} and, thus, leaves a broader window for new physics. This is particularly important in the light of the lithium problem \citep[see., e.\,g.,][]{Fields2011,Coc2014,Cyburt2015}.

\subsection{The baryon density of the Universe}

The (D/H)$_{\rm p}$ value leads to the baryon-to-photon ratio $\eta_{10}$ and, thus, to the baryon density of the Universe $\Omega_{\rm b}h^2$ according to the fitting relation for the BBN calculations (D/H)$_{\rm p}=2.60(1 \pm 0.06)\times 10^{-5}(6/\eta_{\rm D})^{1.6}$ \citep{Steigman2012}, where $\eta_{\rm D}=\eta_{10}-6(S-1)$ with the expansion rate factor $S=(1+7\Delta N_{\nu}/43)^{1/2}$ and extra neutrino species $\Delta N_{\nu}$. The error in the fitting relation stems from uncertainties in nuclear reaction rates, in the D(p,$\gamma)^3$He cross section for the most part \citep[see their discussion in][]{Steigman2012}. The standard BBN ($S=1$ and $\Delta N_{\nu}=0$) yields 
\begin{equation}
\eta_{10} =  6.09\pm 0.36,
\end{equation}
\begin{equation}
\Omega_{\rm b}h^2 = 0.0222\pm0.0013,
\end{equation}
which are in good agreement with the CMB predictions. At the same time, as it was mentioned in the previous section, the confident intervals of $\eta_{10}$ and $\Omega_{\rm b}h^2$ leave a broad window for new physics.

\section{Conclusions}

A new measurement of deuterium abundance in the low-metallicity ($[{\rm O/H}]=-2.042 \pm 0.005$) subDLA system with column density log$N$(H\,{\sc i})~$=19.983\pm0.010$ at $z_{\rm abs}=2.437$ towards QSO J\,1444$+$2919 has been reported. This system has relatively complex velocity structure, thereby the major source of uncertainties in the measured D/H value comes from an assumption of uniform metallicity over the subDLA components. Having estimated this systematics, the value of log\,(D/H)~$=-4.706\pm0.007_{\rm stat}\pm0.067_{\rm syst}$ has been deduced.

This subDLA system meets the set of selection criteria stated recently by \citet{Cooke2014} and, thus, increase the {\it Precision Sample}. However, following other authors we stress importance of relaxing these selection criteria.

As systematic errors in some D/H measurements were likely to be underestimated, weighted mean can be inappropriate to describe the primordial deuterium abundance. Alternatively, unweighted mean can be used for this. An unweighted mean value of the whole D/H data sample of 15 gives a conservative estimate of the primordial deuterium abundance of $({\rm D/H})_{\rm p}=(2.54\pm 0.19)\times10^{-5}$ which is in good agreement with CMB prediction for the standard BBN.

By means of the derived (D/H)$_{\rm p}$ value cosmological parameters such as the baryon-to-photon ratio $\eta=(6.09\pm 0.36)\times10^{-10}$ and the baryon density of the Universe $\Omega_{\rm b}h^2=0.0222\pm0.0013$ has been deduced. Their confidence intervals leave a broad window for standard BBN extensions, what could be extremely important for resolving the lithium problem.

Undoubtedly, precision measurements of deuterium abundance is a powerful tool of studying the Early Universe, verification of the Standard Model and testing new physics. That is why it is vital to widen the sample of unbiased D/H measurements in the nearest future.

\section*{Acknowledgments}

This work has been supported by the Russian Science Foundation (grant No~14-12-00955).
It is also based on observations collected with the Keck Observatory Archive (KOA), which is operated by the W.\,M. Keck Observatory and the NASA Exoplanet Science Institute (NExScI), under contract with the National Aeronautics and Space Administration. 



\bibliographystyle{mnras}
\bibliography{mybibliography} 

\bsp	
\label{lastpage}
\end{document}